\documentclass[%
 reprint,
 amsmath,amssymb,
pre
]{revtex4-2}

\usepackage{graphicx}% Include figure files
\usepackage{dcolumn}% Align table columns on decimal point
\usepackage{bm}% bold math
\usepackage{xcolor}
\usepackage{float}
\begin{document}

\preprint{APS/123-QED}

\title{Rogue ocean waves and the St. Petersburg paradox}

\author{Paul Johns}%
\affiliation{%
 U.S. Naval Research Laboratory\\
 4555 Overlook Ave. Southwest, Washington, DC 20375, USA
}%

\author{Peter Palffy-Muhoray}%
\email{mpalffy@kent.edu}
\affiliation{%
 Advanced Materials and Liquid Crystal Institute\\
 Kent State University, Kent, Ohio 44242, USA
}%

\author{Jake Fontana}%
 \email{jake.fontana@nrl.navy.mil}
\affiliation{%
 U.S. Naval Research Laboratory\\
 4555 Overlook Ave. Southwest, Washington, DC 20375, USA
}%

\date{\today}%

\begin{abstract}
Powerful rogue ocean waves have been objects of fascination for centuries.  Elusive and awe-inspiring, with the potential to inflict catastrophic damage, rogue waves remain unpredictable and imperfectly understood. To gain further insight into their behavior, we analyzed 3,441,188,683 ocean surface waves to determine the statistical height distribution of the largest waves. We found that the distribution of rare events which resolves the St. Petersburg paradox also describes the relative height distribution of the largest waves.  This result is expected to contribute to the modeling of ocean surface dynamics and improve the accuracy of marine weather forecasts.

%\begin{description}
% \item[PACS numbers]
%83.60.Uv, 83.60.-a, 62.25.+g, 83.85.Tz
% \end{description}

\end{abstract}

\maketitle

%\tableofcontents

\section{INTRODUCTION}

Rogue waves are exceptionally large, naturally occurring waves, ranging from ocean surface waves \cite{Zakharov,Dematteis} to waves in optics \cite{Solli}, Bose-Einstein condensates \cite{Bludov}, plasmas \cite{Bailung}, and even finance \cite{Yan}. Although their emergence is infrequent, encounters with them are often catastrophic, underpinning the need to anticipate them. 

Considerable progress has been made over the last 70 years in understanding and forecasting rogue waves.  Yet, ambiguities still exist regarding how well the established models fit observations, particularly when the number of waves becomes large \cite{Solli,Forristall,Steinmeyer, kharif, Christou, Dudley2}. 

To predict rogue waves, an effective description is needed to relate the heights of waves to the number of waves  $N$ in a set.  It is standard practice to monitor the maximum normalized wave height, $H_{max}/H_{s}$, where $H_{max}$ is the maximum wave height and $H_{s}$ is the significant wave height.   The significant wave height $H_{s}$ is the average height of the highest 1/3$^{rd}$ of the waves in the set.   Rogue waves are those whose height $H$ is at least twice the significant wave height, that is, where $H/H_{s}>2$.  Typically, rogue waves appear in sets where $N$ is large and increase in height with increasing $N$.

While many models have been developed to predict $H_{max}/H_{s}$ for ocean surface waves, at least two models have emerged over time which agree well with observations. The Rayleigh model \cite{LONGUETHIGGINS}, developed by Longuet-Higgins, is based on the postulate that rogue waves arise as the sum of contributions from a large number of smaller waves with random phases.   Making use of the Rayleigh distribution, it proposes a square root and logarithm dependence to predict the most probable value, $H_{max}/H_{s}=\alpha \sqrt{ln(N)}+\beta$  where $\alpha$ and $\beta$  are constants. The Weibull model \cite{Forristall}, introduced and empirically justified for ocean waves by Forristall, predicts a generalized combination of a power law and logarithmic dependence, $H_{max}/H_{s}=\alpha (ln(N))^{\beta}$.

Rather than pursuing a detailed model here, we follow the approach of Weibull \cite{Weibull} and Forristall \cite{Forristall} and provide here a simple empirical form which gives an excellent fit to the data.  Based on our analysis of the largest data set reported to date, we demonstrate that the relative heights of ocean surface waves appear to accumulate similarly to profits in the St. Petersburg game \cite{RN1372}. These results may also have applications in other scientific and engineering fields where rogue waves are prevalent.  

In the St. Petersburg game \cite{Bernoulli} a true coin is flipped until it lands heads. If this happens on the nth flip, the player receives $2^{n}$ dollars.   The paradox is that although the expectation value of the payout is infinite, the typical payout is small. A lucid resolution of the paradox, provided by Feller \cite{Feller1945}, is that the length of the longest run of tails, and hence the expected payout per game, depends linearly on the logarithm of the number of flips. 

To form a correspondence between ocean surface wave heights and the game, one can regard ocean waves as the sum of small wavelets.  If the height of waves is taken to be the sum of heights of many small identical wavelets with the same phase, then wave heights accrue similarly to lengths of runs of tails in the St. Petersburg game.  The expected number of waves with dimensionless height $h$ in a set of $N$ waves is $(1-p)^2p^{h}N$, where $p$ is the probability of a wavelet with a given phase. Since there is only one wave with maximum height $h_{max}$ in the set, $(1-p)^2p^{h_{max}}N=1$, and the expected maximum wave height $h_{max}=-ln[N(1-p)^2]/ln(p)$, giving the maximum normalized wave height \cite{Fontana}, 

\begin{equation}
\frac{H_{max}}{H_{s}}=\alpha ln(N)+\beta
\end{equation}

We call this hypothesis the Petersburg model. To verify the proposed relationship between the maximum normalized wave heights and the number of waves by observation, data spanning several orders of magnitude is required due to the logarithmic dependence, as demonstrated in other applications of the Petersburg description \cite{Fontana,Scher,RN1371}.

\section{RESULTS}

The Coastal Data Information Program (CDIP) at U.C. San Diego \cite{CDIP} has recorded 27 years of ocean surface wave data from buoys located predominantly across the Pacific ocean. We analyzed data from 156 buoys from the database, spanning the period August 27, 1993, 16:32 (UTC) $\textendash$ June 30, 2020, 14:27 (UTC), resulting in 3,441,188,683 waves after filtering, the equivalent of a single buoy collecting data for 6$\frac{1}{2}$ centuries (649 years).  

A Python code was used on a desktop computer (128 GB RAM, 6-core 3.4 GHz processors, 32 TB hard drive) to process all the data presented here \cite{SM}. Sensors on the buoys recorded their movement  relative to the still-water line (mean value), produced by the waves as a function of time.  The code used a zero-upcrossing method to identify and measure the height of each wave.  The difference between the maximum and minimum vertical displacements between consecutive zero up-crossings determined the height of each wave.  The data collected by each buoy was filtered to remove spurious events (data gaps, electrical noise, anomalous buoy motion, or wave data flagged by CDIP) using standard quality control methods \cite{orzech}: the wave crest height was restricted to an upper bound of $<1.5\times H_{s}$ and a lower bound of $H_{s}>1$ m, and the wave kurtosis was  restricted to an upper bound of $<6$. Unless otherwise specified, the datasets were analyzed in standard periods of $T=30$ minutes, where $H_{s}$ was determined every $T$ \cite{CDIP}.

An illustrative example of the maximum and significant wave heights per period, $H_{max}^{T}$ and $H_{s}$, is shown in Fig. 1(a), highlighting the largest wave recorded in the CDIP database $(H=25.53~m)$. The maximum versus significant wave heights per period were determined for all the buoy datasets; these are plotted in Fig. 1(b). The data were organized into hexagonal bins (hexbins), where $n$ is the number of waves per bin, displayed as a blue color hue.  The solid black line corresponds to the rogue wave threshold, $H_{max}^{T}/H_{s}=2$, resulting in 615,978 rogue waves. 

\begin{figure}[htbp]
\centering\includegraphics[width=8
cm]{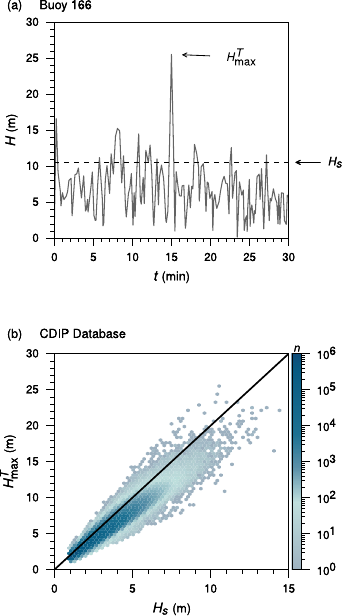}
\vspace{-0.25cm}
\caption{Maximum and significant wave height per period. (a) Wave height evolution for a 30 minute period from buoy 166 (50.017 N, 145.170 W) on December 2, 2016. (b) Hexbin plot of the maximum wave height per period versus significant wave height for all buoy datasets. The solid line is the rogue wave threshold and $n$ is the number of waves per hexbin.}
\label{1}
\end{figure}

The maximum normalized wave height, $H_{max}^{N}/H_{s}$, is the largest $H(N)/H_{s}$ encountered in $N$ waves and is shown for a single buoy (092) in Fig. 2(a) and for all the CDIP buoy datasets in Fig. 2(b).  To determine $H_{max}^{N}/H_{s}$, it is useful to define the normalized wave heights $\hat{H}= H(N)/H_{s}$.  These were calculated initially, updating $H_{s}$ every period. Normalizing  $H(N)$ by $H_{s}$ enables comparison over many time periods, locations, and varying sea states. The normalized waves were then sequentially analyzed per buoy. If  $\hat{H}(N)$ exceeded the heights of all preceding waves, then $\hat{H}(N)$ was used to replace $H_{max}^{N}/H_{s}$.  These results are plotted in Figs. 2(a) and 2(b), showing the maximum normalized wave height a stationary observer (buoy) encounters in $N$ waves.

The Petersburg, Rayleigh, and Weibull models were fit to the data in Figs. 2(a) and 2(b). The adjusted coefficient of determination, $\bar{R}^{2}$, was used as the figure of merit to determine the best fit. As an example for a single buoy (092), with data consisting of over 110,000,000 waves, we find $\bar{R}^{2}$ is 0.946, 0.909 and 0.866 for the Petersburg, Rayleigh, and Weibull models, respectively [Fig. 2(a)]. When data including all of the waves from the CDIP database were examined [Fig. 2(b)], we found that $\bar{R}^{2}$ is 0.892, 0.889, and 0.843 for the three models.

\begin{figure}[htbp]
\centering\includegraphics[width=8.25
cm]{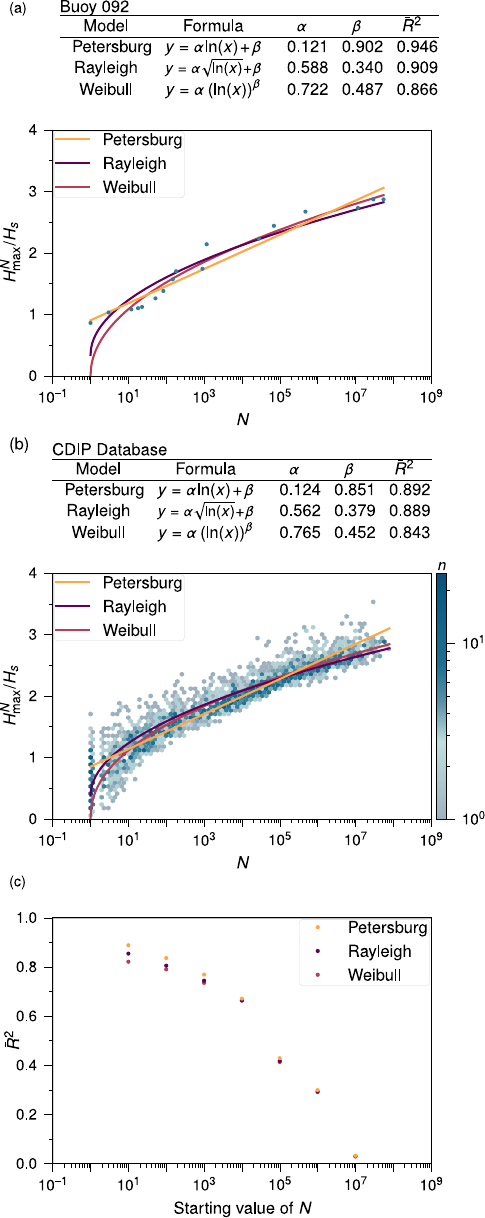}
%\vspace{0.5cm}
\caption{Maximum normalized wave height versus wave number, $N$. (a) Buoy 092 (33.61791 N, -118.31701 E), (b) Hexbin plot for all CDIP buoy datasets, where $n$ is the number of waves per hexbin and (c) the adjusted coefficient of determination for subranges of $N$.}
\label{1}
\end{figure}

\begin{table}[htbp]
\caption{(a) Summary of the individual buoy fits in Fig. 2(b). (b) The effect of varying the period: Petersburg, Rayleigh, and Weibull model fits to   and   for 60 and 15 minute periods over all of the CDIP database.}
\centering\includegraphics[width=8cm]{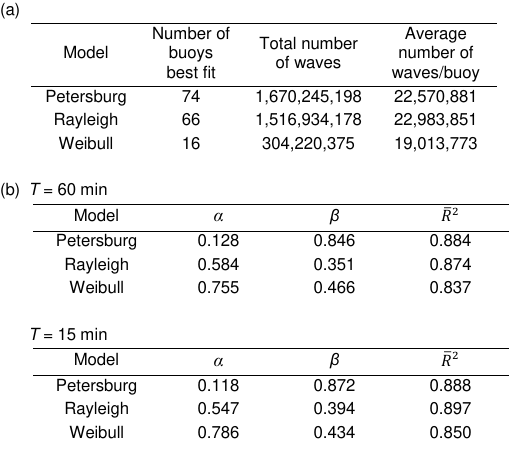}
%\vspace{0.5cm}
\label{1}
\end{table}

The Petersburg model fits the total wave data best [Fig. 2(b)], improving the capability to anticipate the maximum normalized wave height based on the number of waves buoys encounter.  $H_{max}^{N}/H_{s}$ is expected to be less than 1 for the first wave, near unity after ten waves, 2 (rogue) by 10,000 waves and 3 by 30,000,000 waves. Temporally,  $H_{max}^{N}/H_{s}$ will be 1, 2, and 3 after approximately a minute, a day (16 hours), and half a decade, respectively, assuming $3.4\times 10^{9}$ waves/649 years.

The Rayleigh and Weibull models fit the full set of wave data nearly as well as the Petersburg model. It is worth noting that even with 3,441,188,683 measurements there is only a $\Delta \bar{R}^{2}$ of $5.5\%$ between the three models, due to the intrinsic scatter of the data.  Albeit, one important distinction is as $N$ becomes large $>10^{5}$ ($>1$ week), the Rayleigh and Weibull models diverge from the observed data, capturing only the lower limit of the potentially largest and most destructive rogue waves [Fig. 2(b)].  

If the data in Fig. 2(b) is fit for different subranges [Fig. 2(c)], where the starting value of $N$ varies from $10^1$ to $10^7$, then  the Petersburg $\bar{R}^{2}$ remains the best fit for each range.  The $\bar{R}^{2}$ values indicate a poorer fit overall for all three models as the subranges decrease in size.

If the buoy data sets are fit individually [Table 1(a)], as opposed to fitting the entire database [Fig. 2(b)], then the Petersburg model also fits better for a larger number of buoys.

The significant wave height serves as a normalization constant, yet $H_{s}$ depends on the period $T$ as does the relationship between $H_{max}^{N}/H_{s}$ and $N$.  The most commonly used period is 30 minutes \cite{CDIP}, and this was used in our analysis above.  To probe the dependence of the results on the period $T$, the entire CDIP database was analyzed; $T$ was doubled to 60 minutes and halved to 15 minutes with results in Table 1(b). For the $T=60$ minute case, all three models gave a poorer fit overall than in the  $T=30$ minute case, but the relative $\bar{R}^{2}$ rankings did not change. In contrast, when $T$ was halved to 15 minutes, the Rayleigh model fit the data slightly better $(\Delta \bar{R}^{2}=0.9\%)$ than the Petersburg model.  

It appears that as the size of the period $T$ and hence the number of waves per period increases, so does the ratio of the adjusted coefficient of determination $\bar{R}^{2}_{Paradox}/\bar{R}^{2}_{Rayleigh}$ from 0.98 at $T=15$, to 1.00 at $T=30$ minutes and 1.01 at $T=60$  minutes. Distributions which depend on sample size are discussed by Feller \cite{Feller} and Hayer and Andersen \cite{hayer}.  

\section{CONCLUSION}

In addition to providing best fit to the data, the advantage of the Petersburg model is its simplicity. The empirical logarithmic dependence of Eq. (1) is supported by nine decades (3,441,188,683 waves) of oceanographic data.  This result suggests that the relative height of ocean surface waves may accumulate similarly to profits in the St. Petersburg game. Beyond ocean waves, it is interesting to ask whether the Petersburg model will have utility in other research fields where rogue waves are encountered.  An open question is if the Petersburg model will continue to provide the best description for ocean surface waves as $N$ continues to increase, assuming other factors remain constant. Regrettably, indications of climate changes make the latter condition unlikely. For now, according to the analysis and statistical indicators used here, the Petersburg model appears to be the best predictor of the relative heights of ocean surface waves. 

\section{ACKNOWLEDGEMENTs}

This paper was supported by the Office of Naval Research (No. N0001420WX00146, No. N0001421WX00025, and No. N00014-18-1-2624). We thank the Ocean Engineering Research Group (OERG) and Scripps Institution of Oceanography for the maintenance and access to the CDIP database. We also are grateful to Mark D. Orzech and David W. Wang for useful discussions.

\bibliography{references}

\end{document}